\documentstyle[aps,twocolumn,floats]{revtex}
\begin{document}
\newcommand{\bec}{\begin{center}}
\newcommand{\ec}{\end{center}}
\newcommand{\be}{\begin{equation}}
\newcommand{\ee}{\end{equation}}
\newcommand{\beqn}{\begin{eqnarray}}
\newcommand{\eeqn}{\end{eqnarray}}
\newcommand{\bet}{\begin{table}}
\newcommand{\ent}{\end{table}}
\newcommand{\bib}{\bibitem}

\baselineskip 4.2mm

\draft


\title{Optimized Effective Potential Method for Polymers}

\author{P. S\"ule$^1$, S. Kurth$^2$ and V. Van Doren$^1$} 
  \address{$1$ Department Natuurkunde,
RUCA, TSM Group\protect\\ Groenenborgerlaan 171., Antwerpen, Belgium\protect\\ 
 E-mail: sule@ruca.ua.ac.be\protect\\
 $2$ Dept. of Physics and Quantum Theory Group\protect\\
 Tulane University, New Orleans, Louisiana 70118}


\maketitle

\begin{abstract}


The optimized effective potential (OEP) method allows for calculation
of the local, effective single particle potential of density functional
theory for explicitly orbital-dependent approximations to the
exchange-correlation energy functional.
In the present work
the OEP method is used together with the
approximation due to Krieger, Li and Iafrate (KLI).
We present the first application of this method to
 polymers.
KLI calculations have been performed
for the insulating polyethylene and the results have been compared to those
from other orbital-dependent potentials.
Various properties of the band structure are also calculated. The single-particle band gap 
strongly depends on the basis set with larger basis sets yielding narrow gaps.
For certain physical quantities such as the total energy and the exchange energy,
the various orbital-dependent exchange-only and Hartree-Fock results differ only slightly.
For the highest occupied orbital energy the difference is more pronounced.
In order to get the right band gap in OEP the exchange contribution to the derivative discontinuity
is calculated and added to 
the Kohn-Sham gap. The corrected gap obtained by the KLI
approach is $12.8$ eV compared with the Hartree-Fock 
and experimental values of
$16.6$ and $8.8$ eV, respectively.
We observe, however, the strong dependence of the derivative discontinuity
on the basis set.

\end{abstract}

\vspace{1cm}

{\bf 1. Introduction}

 The optimized effective potential (OEP) method  was invented by Talman and Shadwick
\cite{TS}, following the original idea of Sharp and Horton \cite{Sharp}.
In the now classic OEP approach 
a given total energy functional depending explicitly on single-particle orbitals
is minimized under the constraint that these orbitals are solutions of a 
single-particle Schr\"odinger equation with a local, effective potential.
The exchange-only
OEP can be interpreted as the implementation of the exact exchange-only density
funtional theory (DFT) \cite{DFT,OEPreview}.
Very recently OEP is extended to include electron correlation
effects via perturbation theory \cite{GL,Sule3}.
In this theory both the correlation energy and the potential are expanded
using standard perturbation theory. In this way a self-consistent parameter-free 
KS {\em ab initio} scheme can be given. The exact energy functional and the exchange-correlation
potential provided by the Hohenberg-Kohn theorem can be calculated
order-by-order by means of the OEP perturbation theory \cite{Sule3}.

To the best of our knowledge, full OEP calculations have so far only been performed for atoms \cite{OEPreview,KLIatom} and for
semiconductors \cite{Stadele} at the exchange-only level of the theory.
The incorporation of electron correlation effects into the OEP formalism
has been the subject of few studies \cite{OEPreview,Casida}.
Most of these studies, however, have used the complicated self-energy
formalism.

 To reduce the numerical complexity of the OEP method,
 Krieger, Li and Iafrate (KLI) introduced a remarkably accurate analytical
approximation to OEP and transformed the original OEP equations
into an easily manageable form \cite{KLI}.
 Applications to atoms and to solids show that the KLI method yields results
that are nearly identical to those of OEP \cite{KLIatom,LKNI}.
The KLI method has been successfully applied for molecules
\cite{KLImol} and for certain semiconductors \cite{Kotani,Bylander} using pseudopotentials.

  In our previous study we have carried out calculations by means of various
orbital-dependent exchange potentials which are accurate approximations of OEP
and require less computational effort \cite{Sule2}.   
We found that the various orbital dependent potentials provide a narrow 
DFT gap for polyethylene (PE).
KS theory will typically give too narrow valence bands and too compressed 
conduction bands \cite{LKNI}.
The KS theory will only
give an exact result for the highest occupied energy eigenvalue,
therefore, calculated states will in general
not have the desired accuracy as one might hope \cite{DFT,KLI}.
In addition to this,
it has been shown that the difference between the highest occupied
and lowest unoccupied DFT eigenvalues in the $N$-electron system
(which is what is usually called the DFT band gap) is not the true
quasiparticle band gap, but differs from it by the discontinuity in the
exchange-correlation potential when an electron is added to the system \cite{OEPreview,PerdewLevy}.
Therefore,
the so-called derivative discontinuity corrections must be taken into account
in order to get the right gap \cite{OEPreview,Sule2,Gorling}.
Consequently,
it is desirable to calculate the band structure using the
KS method, and then simply add on a single-shot correction to all states \cite{LKNI,Sule2}.
In this way much better agreement is found with the 
Hartree-Fock gap 
when the exchange corrections are considered to the KS eigenvalue band gap 
\cite{Sule2}.
None of the state-of-the-art exchange-correlation functionals depending only explicitly on the density show the abovementioned discontinuity in 
the exchange-correlation potential and therefore do not yield the correct
values for the band gaps of the insulating solids \cite{LKNI}.

In the last decades quite a number of publications appeared on
polymer electronic structure using mainly Hartree-Fock {\em ab initio}
methods \cite{Ladik}. Results are reported only in recent years,
however, using DFT methods like local density approximation (LDA)
\cite{LDApolymer}.
The generalized gradient approximation (GGA) \cite{PW91} resulted in only modest improvement
over LDA gap in semiconductors \cite{Ortiz}.

 In this paper, we briefly review the theory of OEP and KLI methods.
 This work can be taken as the systematic continuation of our previous work
on polymer electronic structure obtained by various orbital dependent
local exchange potentials \cite{Sule2}.
Particular attention is paid to the correction of the OEP eigenvalue band gap.
 We will present the first results obtained by the KLI orbital dependent exchange potential for polymers. 
We give 
 the calculated band gaps, band widths, band structures and total energy of polyethylene
within the Kohn-Sham density functional scheme using various orbital dependent
exchange potentials. 

\vspace{8mm}

{\bf 2. Basic Formalism}\\

 The following spin-restricted Kohn-Sham type mean-field approach is employed for polymers (the extension of the theory for spinpolarized problems is
straightforward):
\be
[-\frac{1}{2} \nabla^2+ v_s({\bf r})] u_i({\bf r})=
 \epsilon_i u_i({\bf r})
\label{kseq}
\ee
where $\{u_i({\bf r})\}$ are the single particle orbitals and
 $v_s({\bf r})$ is the Kohn-Sham effective single particle potential \cite{DFT}.
\be
v_s({\bf r})=v_{ext}({\bf r})+v_H({\bf r})+v_{xc}({\bf r})
\label{kspot}
\ee
The Kohn-Sham one-body potential contains the external, Hartree and exchange-correlation
potential. $v_{ext}({\bf r})$ represents the Coulomb potential of the
nuclei. $v_H({\bf r})$ is the classical Coulomb repulsion of the electron charge density.The exchange-correlation potential $v_{xc}({\bf r})$ is a nonclassical term.
It is formally defined  as functional derivative of the exchange-correlation
energy $E_{xc}[\rho]$:
\be
v_{xc}({\bf r},[\rho])=\frac{\delta E_{xc}[\rho]}{\delta \rho({\bf r})}
\label{funcder}
\ee
 We seek an expression which contains the exact exchange potential
and correlation can be added easily on top of the exact exchange-only
theory using approximate correlation energy functionals \cite{DFT}.
 It must be emphasized, however that the correlation energy functionals
either in their local or gradient corrected form may not work in combination
with exact exchange. 
This is because basically all common correlation functionals do not have
a long-range component in the corresponding correlation hole (the combined
exchange-correlation hole is typically short-ranged).
The Colle-Salvetti gradient corrected correlation functional \cite{CS}, which
provided excellent results for atoms, performed rather badly
for molecules for the abovementioned reason \cite{KLIatom,KLImol}.
Approximate correlation functionals which are derived from the
homogeneous or inhomogeneous electron gas model or obtained non-empirically
from sum-rule conditions \cite{Sule} have an incorrect long-range tail
(only dynamical correlation is accounted for).
The correlation energy functionals provide
improper local behaviour in the bond midpoint regions in molecules \cite{Sule,Gritsenko,Leeuwen}
which is mainly due to the lack of non-dynamical correlation effects
\cite{Sule}.
Therefore, in those methods which work with exact exchange (HF and
OEP) the correlation effect can only be accounted for in a more complicated
way than in the so-called second generation of DFT \cite{Sule2}.

  In the optimized effective potential method (OEP) in contrast to the ordinary DFT, the
exchange-correlation energy is approximated as an explicit functional of orbitals and only 
implicit functional of the electronic density \cite{KLImol}. 
 The starting point of the OEP method is the total energy functional
\beqn
\enspace
E_{tot}^{OEP}[\rho] = \sum_{i=1}^{occ} \int d{\bf r} u_i^*({\bf r}) \biggm(
-\frac{1}{2} \nabla^2 \biggm) u_i({\bf r}) \nonumber \\
+\int d{\bf r} \rho({\bf r}) v_{ext}({\bf r}) 
+ \frac{1}{2} \int d {\bf r} d {\bf r'} \frac{\rho({\bf r}) \rho({\bf r'})}
{\vert \bf{r}-\bf{r}' \vert} 
+ E_{xc}^{OEP}[\{u_i\}].
\eeqn
 We make use of the chain rule for functional derivatives to obtain from Eq.~(\ref{funcder})
\be
v_{xc}^{OEP}({\bf r},[\rho])=
 \sum_{i=1}^{occ} \int d {\bf r}' \frac{\delta E_{xc}^{OEP}[\{u_i\}]}{\delta u_i({\bf r^{'}})}
\frac{\delta u_i({\bf r'})}{\delta \rho({\bf r})} + c.c.
\ee
where formally $\{ u_i [\rho] \}$ are the orbitals which are, however, implicit
functionals of the density.
Applying the functional chain rule again and after some algebra one
obtains the following integral equation:
\be
\sum_{i=1}^{occ} u_i({\bf r}) \int d{\bf r'} [v_{xc}({\bf r'})-v_i({\bf r'})] G_{si}({\bf r},{\bf r'})
u_i^*({\bf r'})+c.c.=0
\label{inteq}
\ee
where
\be
v_i({\bf r})=\frac{1}{u_i^*({\bf r})} \frac{\delta E_{xc}^{OEP}[u_i]}
{u_i({\bf r})}
\ee
and
\be
 G_{si}({\bf r},{\bf r}')=\sum_{k \ne i}^{\infty} \frac{u_k({\bf r})
u_k^*({\bf r}')}{\epsilon_i-\epsilon_k}.
\ee
The integral in Eq.~(\ref{inteq}) is the fundamental expression for $v_{xc}$ 
in OEP and can be shown to be equivalent to Eq.~(\ref{funcder}) \cite{OEPreview}. There is no known analytic
solution for $v_{xc}[\{u_i\}]$. Therefore, only numerical solutions are available
for spherical atoms \cite{TS,OEPreview,KLI,Engel2,Talman} and for solids
\cite{Stadele,Kotani,Bylander} using atomiclike wave functions and assuming
integer atomic occupation numbers within each muffin-tin sphere (linear
augmented plane-wave method) \cite{LKNI}. 
These numerical solutions of Eq.~(\ref{inteq}) are confined to the exchange-only
OEP and correlation has been taken into account only via approximate local
functionals \cite{Stadele}.
In the work of Krieger and co-workers the OEP integral equation is analyzed
and a simple approximation is made which reduces the complexity of the
original OEP equation significantly and at the same time keeps many of
the essential properties of OEP unchanged \cite{KLI}.
Krieger, Li and Iafrate gave an exact expression transforming the OEP
integral equation (\ref{inteq}) into a manageable form. They obtained the following, still  {\em exact}
expression for $v_{xc}({\bf r})$:
\beqn
v_{xc}^{OEP}({\bf r})=v_{xc}^S({\bf r})+\sum_{i=1}^{occ-1} \frac{\rho_{i}({\bf r})}{\rho({\bf r})}
(\overline{v}_{xci}^{OEP}-\overline{v}_i) 
&+& \\ \nonumber \frac{1}{2} \sum_{i=1}^{occ} \frac{\nabla [p_i({\bf r}) \nabla u_i({\bf r})]}
{\rho({\bf r})},
\eeqn
\beqn
 \overline{v}_i=\int d{\bf r} \rho_i({\bf r}) v_i({\bf r}),
\nonumber
\eeqn
where $m=occ$ is the highest occupied one electron energy level.  
$\rho_i=\vert u_i \vert^2$ is the partial density.
$v_x^S({\bf r})$ is the Slater's potential (its correlation part is not known exactly) 
\be
v_x^S({\bf r})=\frac{1}{2 \rho({\bf r})} \int d{\bf r}' \frac{\vert \gamma({\bf r},
{\bf r}') \vert^2}{\vert {\bf r}- {\bf r}' \vert},
\ee
where $\gamma({\bf r},{\bf r}^{'})$ is the first-order density matrix.
The summation runs over the orbital index for all the occupied orbitals up to the
highest occupied $m$th orbital (Fermi level).
The function $p_i$ is defined by 
\be
p_i({\bf r})=\frac{1}{u_i({\bf r})} \int d{\bf r}' [v_{xc}^{OEP}({\bf r}')-
v_i({\bf r}')] G_{si}({\bf r},{\bf r}') u_i({\bf r}').
\ee
In practical applications the last term in Eq. (9) turned out to be quite small
in atomic systems and has small effect only on the atomic shell boundaries \cite{KLI}. 
This additional term's average over the $m$th orbitals is zero \cite{KLI}.
Neglecting this term one obtains the exchange-only KLI-approximation which,
after some algebra, can be written in the following form \cite{KLI}
\be
v_x^{KLI}({\bf r})=v_x^S({\bf r})+
\sum_{i=1}^{occ-1} \frac{\vert u_i({\bf r}) \vert^2}{\rho({\bf r})} \sum_{j=1}^{occ-1} ({\bf A}^{-1})_{ij} (\overline{v}_{xj}^S-\overline{v}
_{xj}).
\label{klieq}
\ee
\be
{\bf A}_{ji}=\delta_{ji}-{\bf M}_{ji},
\ee
\be
 {\bf M}_{ji}=\int \frac{\rho_j({\bf r}) \rho_i({\bf r})}
{\rho({\bf r})} dr, i,j=1,...,m-1.
\ee
$\overline{v}_{xj}^S$ and $\overline{v}_{xj}^{HF}$ can be given as follows:
\be
\overline{v}_{xj}^S= \int \rho_j({\bf r}) v_x^S({\bf r}) dr.
\ee
\be
\overline{v}_{xj} 
 = -\frac{1}{2} \sum_{i=1}^{occ} \int d{\bf r} d{\bf r}' \frac{u_i^*({\bf r})
u_j^*({\bf r^{'}}) u_i({\bf r'}) u_j({\bf r})}
{\vert {\bf r}-{\bf r}' \vert},
\ee
Atomic calculations \cite{KLIatom} show that in all situations, the KLI
potential mimics the OEP results extremely well and even correctly
preserves the property of integer discontinuity \cite{OEPreview,LKNI}.
Moreover, the KLI method is also easy to program.
 In this article we restrict ourself to various exchange-only methods, however,
the extension of these methods to exchange-correlation case is
straightforward using the best available correlation energy functional.

 For the constants $w_i=\overline{v}_{xi}^{OEP}-\overline{v}_i$ (Eq. (9)) an alternative
expression is proposed by Gritsenko {\em et al.} \cite{Gritsenko} in terms of orbital energies
$\epsilon_i$. It follows from gauge invariance requirements, proper scaling and short range behaviour of the response part of Eq. (13) that $w_j$ can
only depend on an energy difference. 
In their work Eq. (13) is approximated by:
\be
 v_x^{SSP}({\bf r})=v_x^S({\bf r})+\frac{8 \sqrt{2}}{3 \pi^2} \sum_{i=1}^{occ-1} 
  \frac{\vert u_i({\bf r})\vert^2}{\rho(
{\bf r})}
 \sqrt{\epsilon_F-\epsilon_i}
\label{ssp}
\ee
We use the notation SSP (Slater's potential + step potential) for this exchange potential.
The parameter
 $\frac{8 \sqrt{2}}{3 \pi^2}$ is determined from the homogeneous electron
gas (HEG) model and therefore Eq.~(\ref{ssp}) is exact in the HEG limit \cite{Gritsenko}.
This constant is chosen as universal parameter for all the calculations.
The main advantage of this expression is that one can avoid the matrix
inversion of Eq. (12).
Note that the summation runs over all the occupied orbitals except the highest one like
in Eq. (12).
With the step potential-like second term $v_x^{SSP}$ provides a good 
approximation to the OEP exchange-potential \cite{Gritsenko}.

  In actual calculations we used the following iterative scheme for getting
$v_x^{KLI}$ instead of Eq.~(\ref{klieq}) in each $k$th SCF cycle
\be
v_x^{KLI(k+1)}({\bf r})=v_x^{S(k)}({\bf r})+\sum_{i=1}^{m-1} \frac{\rho_i^k}{\rho^k}
(\overline{v}_x^{KLI(k)}-\overline{v}_i^{k}).
\label{itkli}
\ee
The initial guess for Eq.~(\ref{itkli}) is chosen as follows
\be
 v_x^{KLI(0)}({\bf r})=\frac{1}{2} (v_x^{LDA}({\bf r})+\frac{2}{3}v_x^S({\bf r})).
\ee
We found this particular form of the initial guess is close to the
KLI exchange potential.

 The KLI approximation to OEP provides a self-interaction free exchange potential and proper $-\frac{1}{r}$ asymptotics as $r \rightarrow \infty$. 
 The approximate potentials $v_x^S$ and $v_x^{SSP}$ exhibit proper asymptotics as well \cite{Gritsenko}. 
LDA or GGA exchange potentials do not show the correct asymptotic
behaviour
\cite{KLI}.
For instance, while the popular GGA exchange-correlation energy functional \cite{PW91}
provides energies for atoms with nice agreement with the exact results, the exchange-correlation
potentials and the energy eigenvalues of the highest occupied state, are both suffering
from error \cite{Sule3,KLI}.
Nevertheless, various forms of GGAs are of great importance and further developments
can certainly be expected in this field in the next future \cite{PK}.


\vspace{8mm} 

 {\bf 3. Computational details}

\vspace{6mm}

Closed shell exchange-only KLI calculations have been performed for polyethylene 
and the results are compared with those from
the Hartree-Fock and Slater approach.
For all the calculations a new code is used which has been developed in our 
laboratory \cite{Diogenes}. 
The various one and two-electronic integrals are taken from
the Erlangen periodic Hartree-Fock program \cite{Erlangen}. 
During the calculations the number of $k$-points is set to $25$. 
The SCF iterations are continued until the density matrix elements changed
by less then 
$10^{-5}$.
$876$ and $778$ grid points are used for heavy atoms and for hydrogen in the numerical
integration using Becke's fuzzy cell method \cite{Becke}.
Thus the total number of grid points per unit cell is $2432$.
Further computational details of our implementation of polymeric calculations
are given elsewhere \cite{Sule2}.
  For polyethylene the experimental "zig-zag" geometry is used \cite{LDApolymer}.
Since the various properties of polyethyene depend significantly on the geometry, we give
the structure we have used for our calculations: the C-C and C-H bond lengths are set to
$1.54$ and $1.10 \AA$ while for the CCC, HCH and HCC angles $113$, $108$ and $113$ degrees are used.
Recent geometry optimization calculations with various DFT functionals provided values
in very nice agreement with these experimental bonding parameters \cite{Hirata}.
Two types of bases are used: the Clementi's minimal (7S/3P) and the larger double-$\zeta$ (9S/5P)  basis sets
 for carbon and the 4S and 6S for hydrogen atom, respectively  \cite{Clementi}.
The (11S/7P) basis set is also employed in certain cases.
  To make comparison with previously published atomic and molecular results \cite{KLIatom,KLImol} and also to check
the reliability of our code we have made calculations in the atomic
and molecular limit. To do so, we set the translation vector to roughly $20.0$ a.u. and choose
the interaction only between the nearest-neighbour $CH_2$ unit cells \cite{Ladik}. Using these parameters the computed
properties must be very close to those obtained by atomic or molecular codes.
The test provides nice agreement with atomic calculations obtained by Hartree-Fock,
Slater's, KLI or SSP exchange-only method which confirms the reliability of our code.
Further computational details of our implementation of KLI for polymers are
given elsewhere \cite{Sule2}.

\vspace{6mm}

{\bf 4. Results}\\

\vspace{3mm}

The comparison is carried out for Be, LiH and for $N_2$ \cite{KLImol,Gritsenko}
in the molecular limit and the energetical results are summarized in Table I.
For the total energy we obtained $-14.569$, $-14.562$, $-14.568$ a.u. for Be using
HF, Slater and KLI methods (with Clementi's minimal basis set). The deviation
from HF can be compared with results obtained by others \cite{Gritsenko}:
$7$ and $1$ mHartree for Slater and KLI, while Gritsenko {\em et al.} obtained 11 and 0.0, 
respectively.
For LiH the corresponding numbers are: $-7.9543$, $-7.9542$, $-7.9542$ a.u. The deviations: 
0.0 , 0.0 mH, while Grabo {\em et al.} \cite{KLImol} found 6 mH for Slater.
The discrepancy between our and other results can be considered rather small and is
probably
due to the different basis set applied here. 
For LiH the highest occupied orbital energies are as follows: $-.2914$, $-.3228$ and $-.3227$ a.u.
which are compared with Grabo's values \cite{KLImol}, $-.3017$ (HF), $-.3150$ (Slater) and $-.3011$ (KLI).
 Finally we give the test of our code for $H_2$ as well. For two-electronic systems
the results obtained by HF or KLI must be identical \cite{KLImol}. For $E_{tot}$ and $\epsilon_m$
(total energy and highest occupied energy level)
the values $-1.12796$,$-1.127953$ and $-.5948$,$-.5913$ a.u. are obtained by HF and
KLI, respectively. The small difference must be attributed to numerical
inaccuracy in the numerical integration (KLI).
Results are also given for $N_2$ in Table I.
 The KLI total energy for $N_2$ differs from the HF one by $9$ mH which is
comparable with that obtained by a fully-numerical basis-set-free code \cite{KLImol}.

 Our results for some physical properties of polyethylene are shown in Tables II-III. 
The results are listed in terms of the order of the neighbouring interaction.
Actually, it turns out that at least 5 neighbours are necessary in order to get convergent
results for all the physical properties we are interested in within the finite neighbouring approach \cite{Sule2}. For comparison
we have calculated all the properties at higher number of neighbouring cells
\cite{Sule2}

  Our calculation using the large (11S/7P) basis set provides a deeper Hartree-Fock total
energy ($-39.02326$ a.u.) than the one obtained by Suhai \cite{Suhai} using the 6-31G* basis set
($-39.02251$ a.u.).
According to the expectations
 the comparison of total energies shows us that Hartree-Fock provides the deepest
energies while the various Kohn-Sham schemes result in somewhat
higher energies. 
In principle the total energy is always lower in HF than in (exchange-only) OEP
since the use of a local potential imposes one more variational contraint on the orbitals \cite{OEPreview}.
In general, the following inequality holds: $E_{HF} < E_{OEP} < E_{KLI} < E_{LDA}$
\cite{KLI2}. KLI is always an upper bound to OEP and always lower
than $E_{LSDX}$ (local exchange-only spin-density theory). 
The difference is around $10$ mH for KLI. For molecules T. Grabo and
E. K. U. Gross \cite{KLImol} found differences around $7$-$8$ mH for KLI.
For a larger basis set KLI and HF total energies differ from each other more significantly ($23$ mH), which
is much smaller, however,  than the corresponding difference for Slater ($44$ mH).
We believe that the bulk part of the difference is not due to the approximate
nature of the exchange potentials applied here, but to the different nature of Hartree-Fock
and DFT approaches in accordance with earlier studies \cite{KLI,KLImol}.
These results must be compared with the values obtained by calculations in the
molecular limit \cite{KLImol}. Treating the $CH_2$ unit cell in the molecular limit (Table I),
one can see that all the properties (total energy, one electron energies) are 
close to the Hartree-Fock values. However, in polymer calculations we got for the
one electron energies quite significant deviation from the Hartree-Fock values (Tables II-III): 
employment of local exchange potentials generally leads
to higher Fermi level and and at the same time to lower first virtual levels. Therefore,
the single particle band gap of polyethylene is smaller for the KS methods than for HF.

 We also observe the peculiar behaviour of
the highest occupied orbital energies $\epsilon_m$ obtained by KLI, which differ from HF values significantly
and are closer to $\epsilon_m^{X\alpha}$ \cite{Sule2}.
This latter result is rather surprising since atomic and molecular calculations show that
the $\epsilon_m$ values are very close to those obtained by HF or OEP \cite{KLI,KLIatom,KLImol}.
Also, in HF we found no big difference between the highest occupied orbital energy in
the molecular limit and in the infinite system.
While this comes as a surprise we attribute the increase of the Fermi level due 
to periodic effects, which come into play only, when one employs localized exchange-potentials for
infinite systems.
One can follow how the periodic effect comes into play when the change of $\epsilon_m$ 
is examined in terms of the order of the neighbouring interaction in Table II-III.
As may be read from Table II-III., when $neig=1$, $\epsilon_m$ obtained by Slater or KLI 
is the lowest one, however, when $neig$ increased the Fermi level is
getting closer to zero progressively. 
 Hartree-Fock does not produce such a
phenomenon since it has a {\em non-local} orbital dependent exchange-potential, which is always deeper than the local counterpart,  and thus
is keeping the Fermi level at a deeper energy level. 
However, using the largest basis set in this study (11S/7P) we obtained
somewhat higher $\epsilon_m^{HF}$ (Table III) as well.
As may be seen from Table III, by increasing the number of neighbours and the
basis set $\epsilon_m^{KLI}$ moves upward significantly and becomes higher than the
corresponding LDA value of $-0.242$ a.u. \cite{Sule2}, except for
SSP.
As can be seen in Tables III,
increasing the basis set reduces the calculated KS gap for all the methods, which is
mainly due to the lowering of the first virtual level $\epsilon_{m+1}$ and to the upward
movement of $\epsilon_m$.
While the HF single-particle gap is still too wide ($16.6$ eV), the calculated Slater's and KLI eigenvalue gaps are too narrow ($3.4$ and $5.1$ eV). 
Others reported the value of $16.7$ eV for the HF gap quite close to our value
\cite{Bartlett}. It is worth to note, however, that other calculations with 6-31$G^{**}$ 
basis set resulted in the smaller $13.4$ eV for the HF gap \cite{Suhai}. Recent reports proved that the
 band gap strongly depends not only on the basis set but also on the quality of the
finite neigbour approach \cite{Bartlett}.

 It is known 
 \cite{PPLB} that the exact exchange-correlation potential
exhibits a discontinous jump as the number of electrons passes through
an integer. This derivative discontinuity plays an important role
in the calculations of the exact band gap which can be written as
\beqn
 \Delta=\Delta_{nonint}^{KS}+\Delta_{xc}~~~~~~~~~~~~~~~~~~~~ \\ \nonumber 
 = \epsilon_{N+1}^{KS}(N)-\epsilon_N^{KS}(N)+\Delta_{xc}~~~~,
\eeqn
where $N$ is the number of electrons, $\Delta_{KS}^{nonint}$ is the Kohn-Sham band gap and $\Delta_{xc}$ is the derivative discontinuity.  
For continuum approximations like LDA or GGA this discontinuity vanishes. However,
in OEP as well as in our (orbital-dependent) approximations to OEP,
the discontinuity is finite \cite{OEPreview,Sule2,Gorling}.
The exchange-only contribution to the band gap, $\Delta_x$, is given by \cite{Stadele,Gorling}
\beqn
\Delta_x(i \rightarrow \nu)=\langle u_{\nu} \vert \hat{v}_x^{HF}-\hat{v}_x^{OEP}[\rho] \vert u_{\nu} \rangle 
 \\ \nonumber 
 - \langle u_i \vert \hat{v}_x^{HF}-\hat{v}_x^{OEP}[\rho] \vert u_i \rangle
\\ \nonumber
-\langle \nu i \vert \nu i \rangle-2 \langle \nu i \vert i \nu \rangle,~~~~~~~
\eeqn
with $\hat{v}_x^{HF}$ being the nonlocal HF exchange operator, however, constructed
from the $N/2$ occupied KS orbitals.
In our previous study \cite{Sule2}
 we have calculated $\Delta_x$ using Slater and SSP methods and the results can also 
be seen in Table II-III. 
The last two terms on the r.h.s. of Eq. (20) will vanish for systems with 
periodic boundary conditions and an infinite number of unit cells \cite{Stadele}.
Since we only take a finite number of cells into account, we will also get
contributions to $\Delta_x$ from these terms. We have studied the convergence
of $\Delta_x$ and, for a given basis set, at $neig=5$ the discontinuity seems
to be poorly converged with respect to the basis set and the results have to be
read with due caution.
 In principle
the discontinuity is guaranteed to be smaller than the true band gap by
its definition \cite{OEPreview}. However, calculations
do not confirm this expectation for semiconductors \cite{Stadele}. 
Nevertheless, $\Delta_x$ brings the KS gap somewhat closer to the HF one. 
With the minimal basis set $\Delta_x$ represents
a rather small correction to the band gap and actually is negative (Table II).
This is in contrast with the finding that $\Delta_x$
amounts to typically twice the band gap \cite{Stadele}.
However, the situation changes completely when a larger basis set is used
(Table III).
The corrected band gap obtained
by the larger basis set with the KLI method (Table III) is $12.8$ eV 
while the Slater and SSP approach yield $9.7$ and $12.7$ eV, respectively
\cite{Sule2}.  

 The much more involved second-order self-energy calculations resulted in $10.3$ eV for the quasi-particle (QP) band gap \cite{Ladik,Suhai}.
Further corrections to $\Delta$ would certainly result in a gap close
to the QP and experimental values.
According to the expectations
  KLI and SSP provide values not far from the HF one while the naked Slater potential
yields somewhat higher band gap then the experimental value of $8.8$ eV 
\cite{Ladik,Exp}.
In the basis set limit one expects the KLI gap to be close to the 
HF value.
The discrepancy beetwen the KLI gap and the experimental one must be partly attributed to the
correlation contribution to the derivative discontinuity ($\Delta_c$).
 It is also useful to make comparison with quasi-particle calculations
where one corrects the HF single-particle gap to get the quasi-particle
band structure \cite{Ladik,Bartlett,Suhai}.
According to the results of Suhai obtained for polyethylene \cite{Suhai}
 the QP correction reduced the HF gap from $15.0$ eV to $10.3$ eV
while the derivative discontinuity correction raises the OEP 
gap from $5.1$ eV up to $12.8$ eV. It would be rather interesting to add
on further corrections to $\Delta_x$ including correlation at second order
level of theory \cite{Gorling}. However, these kind of calculations  
are beyond the scope of the present article.

 Atomic calculations provided KLI eigenvalues which are very close to the
HF and OEP values \cite{KLIatom}.
 In molecules, however, 
  Grabo {\em et al.} found slight deviations from the HF values \cite{KLImol}.
 Approximate OEP calculations on polyethylene also resulted in eigenvalues
different from HF \cite{Sule2}.
 In Table II-III we have already demonstrated that this is true 
in the KLI approach as well.
 In order to make a more detailed analysis of this problem
  we give the one-electron energies obtained by HF and KLI
exchange-only methods in Table IV.
We compare the HF and KLI eigenvalues obtained in the molecular limit and in the polymer.
In general HF provides deeper eigenvalues by $100$ to $200$ mH for the core orbitals
and much less difference for the valence orbitals. 
The reverse is true for the KLI virtual orbital energies.
The KLI $\epsilon_m$ value of $-0.093$ eV is compared with the
values of $-0.379$ and $-0.242$ eV obtained by HF and LDA \cite{Sule2}, respectively.
To sum up, the KLI eigenvalues shift upward in the valence region and
shift downward in the virtual space when compared with the HF values in
the polymer.
The KLI highest occupied orbital energy is rather close to the HF one in the
molecular limit. The differences are around $15$ and $100$ mH, in the molecule and in the polymer.
 
  The exchange energies are also presented for various methods. 
 KLI produces a higher exchange energy than HF by some $30$ mH. 
The reason for this has to be found
in the different form of the virial theorem which holds for HF and for DFT.
In DFT the virial theorem is as follows \cite{LP},
\be
  E_x[\rho]=\int d{\bf r} [3 \rho({\bf r})+{\bf r} \nabla\rho({\bf r}) ] v_x({\bf r}),
\label{oepex}.
\ee
 On the basis of Eq.~(\ref{oepex}) 
  deeper exchange-potential will provide deeper exchange energy for the same
density.
 Among the model potentials employed in this study the Slater potential
exhibits the deepest potential curve in atoms \cite{KLI,LKNI} and in Table III one can see
that the exchange energy $E_x^{Slater}$ is the deepest and differ from Hartree-Fock
by some $50$ mH. KLI provides exchange potential similar to that of LDA \cite{Gritsenko,LKNI} in magnitude, although the shape of the curves are different (see e.g. in refs. \cite{KLI,LKNI}). The Fermi level $\epsilon_m$ is deepest for the Slater's
potential among the KS-based methods due to the deepest exchange potential.
Even the first virtual level dropped below zero as obtained by the largest
basis set by the Slater's exchange potential.
 The exchange energy obtained by this potential is approximately by $1 \%$
lower than the Hartree-Fock exchange energy. This difference is around the magnitude of
the non-dynamical correlation energy \cite{Sule}. Therefore the Slater's
potential is a 
suitable candidate for further calculations together with
the commonly used correlation energy functionals which account for only the
short range type dynamical correlation energy.
The success of the popular exchange energy functionals, such as
the one of Becke \cite{Becke} and the GGA's \cite{PW91} lie in the overestimation of Hartree-Fock
exchange and therefore in the effective treatment of exchange-correlation
mainly due to the cancellation of errors with opposite signs \cite{Sule}.
  On the basis of Table III $v_x^{SSP}$ provides the best agreement 
either with the Hartree-Fock total energy or with the exchange energy.
 Therefore one might speculate that the SSP potential is somewhat closer to
the exact OEP potential than to the KLI potential.  
The $v_x^{SSP}$ is the functional not only of the orbitals but also
of the eigenvalues 
of the KS problem while $v_x^{KLI}$ is only of the orbitals.
  Further steps in this direction can be desirable in the future
in developing a general and accurate potential $v_{xc}^{OEP}[\{
u_i,\epsilon_i\}]$ \cite{Sule3} which is a functional of the
eigensolutions of the KS problem.

 In Fig. 1 the band structure is plotted obtained by various methods using
the Clementi's double-$\zeta$ basis.
One can see that virtual
bands are appearing below the zero energy level obtained only by the Slater potential. 
Note that the first three curves in both of the figures are the occupied valence energy levels 
(Fig. 1 and 2, respectively, the lowest, core level is not plotted). 
We also note from Fig. 1 that
the highest three virtual levels are similar to the virtuals
obtained by the minimal basis set (see Fig. 1 in ref. \cite{Sule2}).
However, new patterns appear for the first two virtual levels as obtained
by the larger basis set (plotted with dashed lines on Fig. 1).
The lowest transition from the Fermi level to the first virtual is appearing
at the edge of the Brillouin zone ($k=1$). However, around $k=0.5$ one can 
see quasi-degeneracy of certain bands or even crossing of virtual levels
close to the edge of the Brillouin zone.

  The maximum of the valence band is appearing in the range of [-2,-11] eV with
the lowest and highest values for
KLI and Hartree-Fock, respectively, compared to the experimentally suggested
values of ionization potential of $9.6-9.8$ eV \cite{Exp}.
Others give lower experimental values in
the $7.6-8.8$ eV range \cite{Ladik,Springbord}.
KLI provides the much lower $2.3$ eV value  while Slater and SSP
give $7.6$ and $4.4$ eV \cite{Sule2}, respectively.
By density functional linear muffin-tin orbital method the value of $5.1$ eV is obtained
for helical polyethylene \cite{Springbord}.
The calculated lowest valence band widths for $X\alpha$, HF, SSP, Slater's and KLI approach are
$6.2$, $9.0$, $4.0$, $3.6$ and $3.9$ eV respectively, which are to be compared with the
experimental $7.2$ eV \cite{Exp}.
The total valence bandwidths are $14.0$ eV ($X\alpha$), $15.4$ eV (SSP), 
$15.6$ eV (KLI) and
$19.8$ eV (HF), compared to a experimental value of $16.2$ eV \cite{Exp}.
Slater, KLI and SSP perform quite well for this band width. 
The calculated gap between the lowest valence band and the minimum of the
higher valence bands is $3.3$ eV (HF), $1.9$ eV ($X\alpha$), $3.0$ (SSP), $3.2$ (Slater) \cite{Sule2} and $2.8$ eV (KLI)
compared with the experimental 2.0 eV \cite{Exp}. 
The HF and $X\alpha$ values are in accordance with those obtained by others
\cite{Ladik,LDApolymer}. 
The bottom of the valence band with $\sigma$ symmetry is $18.0$ eV obtained
by KLI which is compared with the HF ($31.0$ eV) and Slater ($23.6$) eV values.
 The overall conclusion can be drawn that the various orbital dependent potentials provide
band structures similar to one another and all of them differ from 
HF. The largest deviation from experiment is found for the lowest
valence band width (LVBW) which is rather narrow in the OEP methods.
 HF give too wide LVBW while X$\alpha$-LDA yields a value in closest agreement
with experiment.


\vspace{2cm}

 {\bf Conclusions}\\

 Density functionals of the third generation treat both the kinetic and
exchange energy exactly at the orbital dependent level of the theory. These
orbitals come from a local potential and, due to the Hohenberg-Kohn theorem,
are therefore functionals of 
the density.
 The optimized effective potential method, in the form given by Talman and
Shadwick \cite{TS} allows one to treat orbital-dependent functionals
in the framework of Kohn-Sham density functional theory.
 The OEP integral equations are difficult to solve for extended systems
and therefore several approximations are introduced. Among them the
one obtained by Krieger {\em et al.} (KLI) and another one derived by Gritsenko {\em et al.}
have become known in the last few years.

 In this work we have applied the exchange-only KLI approximation to the OEP approach
for insulating polymers.
 Calculations have been carried out for polyethylene with a new polymer code using the efficient KLI approximation to OEP.
The results obtained by
the "naked" Slater's potential 
and KLI 
are compared with those obtained by Hartree-Fock. 
The addition of the step-potential as a response part of the exact exchange (Eq. 13) to the Slater's
potential considerably improves the quality of various physical properties calculated
for polyethylene.
  To test the quality and reliability of our code we performed calculations
in the atomic and molecular limit. The results can be brought into harmony with those
obtained by others.

 In general, we find that the band structure calculated by different $v_x$ are 
similar with little qualitative difference, while all of them differ significantly from
HF which have valence bands much deeper and much higher virtual levels.
 The single-particle band gap obtained by the KLI or Slater approaches is too narrow
when compared with the experiment.
The shifted values are observed for the Fermi level, which differ from Hartree-Fock values
significantly. 
 The one electron energy levels are carefully examined and an upward shift of
eigenvalues observed when compared with those of HF, and also the counter
shift of virtual levels. The deviation from the HF eigenvalues
is more significant then in molecules or in atoms.
In order to make a valuable comparison with Hartree-Fock, LDA and with the
experimental gap, correction must be taken into account to the
Kohn-Sham eigenvalue gap.
 The exchange contribution to the derivative discontinuity is calculated
by the KLI exchange potential.
In this way the KS single particle gap is corrected by a single shot 
calculation on top of exchange-only OEP.
 The exchange contribution to the band gap is somewhat less than twice the KS eigenvalue gap.
According to the expectations
the corrected gap is much closer to the Hartree-Fock value than the
KS eigenvalue gap.

\bec

\baselineskip 5mm 

\vspace{8mm}

{\scriptsize
{\bf Acknowledgments}}

\ec

{\scriptsize
This work was supported by the Flemish Science Foundation.

\vspace{1cm}





\newpage

\begin{table}
\caption[]{Comparison of our HF  and KLI results (Clementi's double zeta basis) 
using the code DIOGENES \cite{Diogenes} with values from the literature.}
{\small $E_{tot}$ and $\epsilon_m$ are the total energy and the highest occupied
orbital energy in a.u.}\\
\scriptsize
\begin{tabular}{llcccc} \hline
    &    & HF$^{our}$ & HF$^{other}$ & KLI$^{our}$ &
           KLI$^{other}$ \\ \hline 
Be  & $E_{tot}$ & -14.569 & -14.5730 \tablenotemark[1] &  
 -14.5684 & -14.5723 \tablenotemark[2] \\  
    & $\epsilon_m$ & -0.3089 & -0.3093 & -.3092 & -.3089 \\ \hline
LiH & $E_{tot}$ & -7.954 & -7.987 \tablenotemark[2] &  
 -7.954 & -7.987 \tablenotemark[3] \\
  & &   -7.8629 \tablenotemark[4] & -7.8620 \tablenotemark[5] & & \\
    & $\epsilon_m$ & -0.2914 & -0.3017 \tablenotemark[2] &   
 -0.3227 & -0.3011 \tablenotemark[3] \\
 & & -.2870 \tablenotemark[4] & -.2857 \tablenotemark[5] & & \\ \hline 
 $N_2$ & $E_{tot}$ & -108.892 \tablenotemark[6] & -108.994 \tablenotemark[4] & -108.883 \tablenotemark[6] & -108.986 \tablenotemark[3] \\ 
   & $\epsilon_m$. & -0.6352 \tablenotemark[6] & -0.6152 \tablenotemark[3] & -0.6659 \tablenotemark[6] & -0.6818 \tablenotemark[3] \\ \hline 
$CH_2$  & $E_{tot}$ & -38.8650 \tablenotemark[6] & & -38.8632 \tablenotemark[6]& \\ 
   & $\epsilon_m$ & -0.3750 \tablenotemark[6] & & -0.3595 \tablenotemark[6] &  \\ \hline 
\end{tabular}
\tablenotemark[1] \cite{KLI}\\
\tablenotemark[2]\cite{KLIatom}\\
\tablenotemark[3] \cite{KLImol}\\
\tablenotemark[4]{with STO-3G basis}\\
\tablenotemark[5]{with STO-3G basis using Gaussian94 \cite{G94} program package}
\\
\tablenotemark[6] The large 11S/7P basis set is used for $N_2$ and for $CH_2$.
\end{table}

\newpage

\begin{table}
\caption[]{Calculated properties of polyethylene by various DFT methods using the 
minimal basis set\\
{\small
 HF, Slater, SSP and KLI denote the Hartree-Fock,
exchange-only method with Slater's potential, Slater's potential with orbital dependent step potential and the KLI exchange-only OEP method.
 All the properties are in a.u. except the HOMO-LUMO
gap which is given in eV. 
The Clementi's minimal basis set (7S/3P) is used.
{\em Neig} gives the number of neighbouring cells taken into account.
$E_{tot}$, $E_x$ are the calculated total and exchange
energies per $CH_2$ unit cell.
$\Delta_x$
is the exchange component of the derivative discontinuity. 
The calculations are carried out for the $CH_2$ unit cell.
}}

\begin{tabular}{lclccr}  \hline 
 neig  &  & HF & Slater \tablenotemark[1] & SSP \tablenotemark[1] & KLI      \\ \hline 
 1 & $E_{tot}$ (a.u.) &  -39.03535 & -39.01888 & -39.02768 & -39.02042   \\ 
 & $E_x$ (a.u.)         & -5.82128   & -5.84774  & -5.79080  & -5.7713  \\
 & $\epsilon_m$ (a.u.)  & -.6672     &  -.5634   & -.4924    & -.3524 \\
 & $\epsilon_{m+1}$ (a.u.) & .3978   &  .0074   & -.1084    & .2228 \\ 
 & {\em gap} (eV)       & 29.0       & 15.1      & 16.4      & 15.7 \\  
 & $\Delta_x$ (eV)      &            & -5.7      & -4.4      & -3.2 \\ \hline
 5 &                    & -38.88012  & -38.86891 & -38.87278 & -38.86981 \\
   &                    & -5.75048   & -5.76397  & -5.7379   & -5.7297 \\
   &                    & -.4747     & -.3524    & -.2312    & -.0871  \\
   &                    & .3444      & -.0433    & .1310     & .2786 \\ 
   &                    & 22.3       & 8.4       & 9.9       & 10.0 \\ 
   &                    &            & -3.5      & -1.9      & -1.2 \\ \hline
 8 &                    & -38.88008  & -38.86890 & -38.87275 & -38.86954 \\
   &                    & -5.75049   & -5.76400  & -5.7378   & -5.7284 \\
   &                    & -.4760     & -.3518    & -.2318    & -.0972 \\
   &                    & .3436      & -.0425    & .1303     & .275 \\ 
   &                    & 22.3       &  8.4      & 9.9       & 10.1 \\ 
   &                    &            & -3.5      & -1.9      & -1.1 \\ \hline
   &  KS gap$+\Delta_x$ &            &  4.9      &  8.0      &  9.0  \\ \hline
\end{tabular}
\tablenotemark[1] \cite{Sule2}
\end{table}
 
\newpage

\begin{table}
\caption[]{Calculated properties of polyethylene by various DFT methods using the 
Clementi's double-zeta basis set (9S/5P).\\
{\small Results with basis (11S/7P) are also given in the bottom of the Table.
 The sum of the calculated KS eigenvalue gap $\Delta_{nonint}^{KS}$ and the
exchange derivative discontinuity $\Delta_x$ is also given together with the exchange-only LDA and experimental
gap (eV).
The notations are the same as in Table I.\\
}}

\begin{tabular}{lclccr}  \hline 
 neig  &  & HF & Slater \tablenotemark[1] & SSP \tablenotemark[1] &  KLI \\ \hline 
 5 & $E_{tot}$    &  -39.01068 & -38.96683 & -38.99704 &  -38.99311    \\ 
 & $E_x$            & -5.8783    & -5.9368   & -5.8841   & -5.8405    \\
 & $\epsilon_m$     & -.4049     & -.2951    & -.1628    & -.1023  \\
 & $\epsilon_{m+1}$ & .1489      & -.1590    & -.0007    & .0284   \\ 
 & KS gap           &  15.1      & 3.7       & 4.4       & 3.6     \\  
 & $\Delta_x$       &            & 3.4       & 3.8       & 1.9      \\ \hline
 8 & (9S/5P)        & -39.01056 & -38.96684  & -38.99760 &  -38.99407    \\
   &                & -5.8780   & -5.9367    & -5.9036  & -5.8621   \\
   &                & -.3984    & -.2774     & -.1130   & -.0997    \\
   &                & .1558     & -.1655     & -.0755   & .0322  \\ 
   &                & 15.1      & 3.7        & 4.4      & 3.6   \\ 
   &                &           & 3.2        & 3.7      & 2.4  \\ \hline
 8 &  (11S/7P)      & -39.02326 &  -38.99926  & -39.01361 & -39.00701    \\
   &                & -5.8977   & -5.9486     & -5.9036 &  -5.8553       \\
   &                & -.3786    & -.2401      & -.1130 & -.0933 \\ 
   &                & .2320     & -.1151      & .0755  & .0958  \\
   &                & 16.6      & 3.4         & 5.1     & 5.1 \\
   &                &           & 6.3         & 7.6    &  7.7  \\ \hline  
   &  KS gap$+\Delta_x$  &      & 9.7         & 12.7   & 12.8  \\ \hline
   &  LDA           &           &             &       & 7.6 \tablenotemark[1] \\  
   &  EXP           &           &             &       & $8.8$ \tablenotemark[2] \\ \hline
\end{tabular}
 \tablenotemark[1] \cite{Sule2}\\
 \tablenotemark[2] \cite{Ladik,Exp}\\
\end{table}

\newpage

\vspace{1cm}
\begin{table}
\caption[]{One-electron energies (a.u.) obtained by HF and KLI exchange-only
approaches using the Clementi's (11S/7P) basis set and $8$
neighbours for polyethylene ($k=1$)}
\begin{tabular}{lcccr}  \hline 
   & ${\bf HF}^{mol}$ & ${\bf KLI}^{mol}$  & ${\bf HF}^{poly}$ & ${\bf KLI}^{poly}$  \\ \hline
  $\epsilon_1$  & -11.289 & -10.201 & -11.219 & -9.820 \\
  $\epsilon_2$  & -.904 & -.774 & -.800 & -.511 \\
  $\epsilon_3$  & -.587 & -.537 & -.681 & -.401  \\
  $\epsilon_4$  & -.375 & -.359 & -.379 & -.093  \\ \hline
  $\epsilon_5$  & .056 & -.318 &  .232 & .096  \\
  $\epsilon_6$  & .240 & -.058 &  .232 & .109  \\
  $\epsilon_7$  & .314 & -.005 &  .392 & .301  \\
  $\epsilon_8$  & .518 & .192 & .719 & .527  \\
\end{tabular}

\end{table}

{\bf Fig. 1} The calculated valence band structure obtained by
various exchange-only methods as a function of the dimensionless variable $k$ with $k=0$ being the zone center and $k=1$ the zone boundary. The Brillouin zone is that corresponding to a $CH_2$ unit cell. Solid lines correspond to
the occupied levels while the dashed lines to the virtual levels.
The two lowest dashed curves of virtual levels are of particular
interest (see text).
The Clementi's double-$\zeta$ basis set is used and 8 neighbours are
considered for the $CH_2$ unit cell.

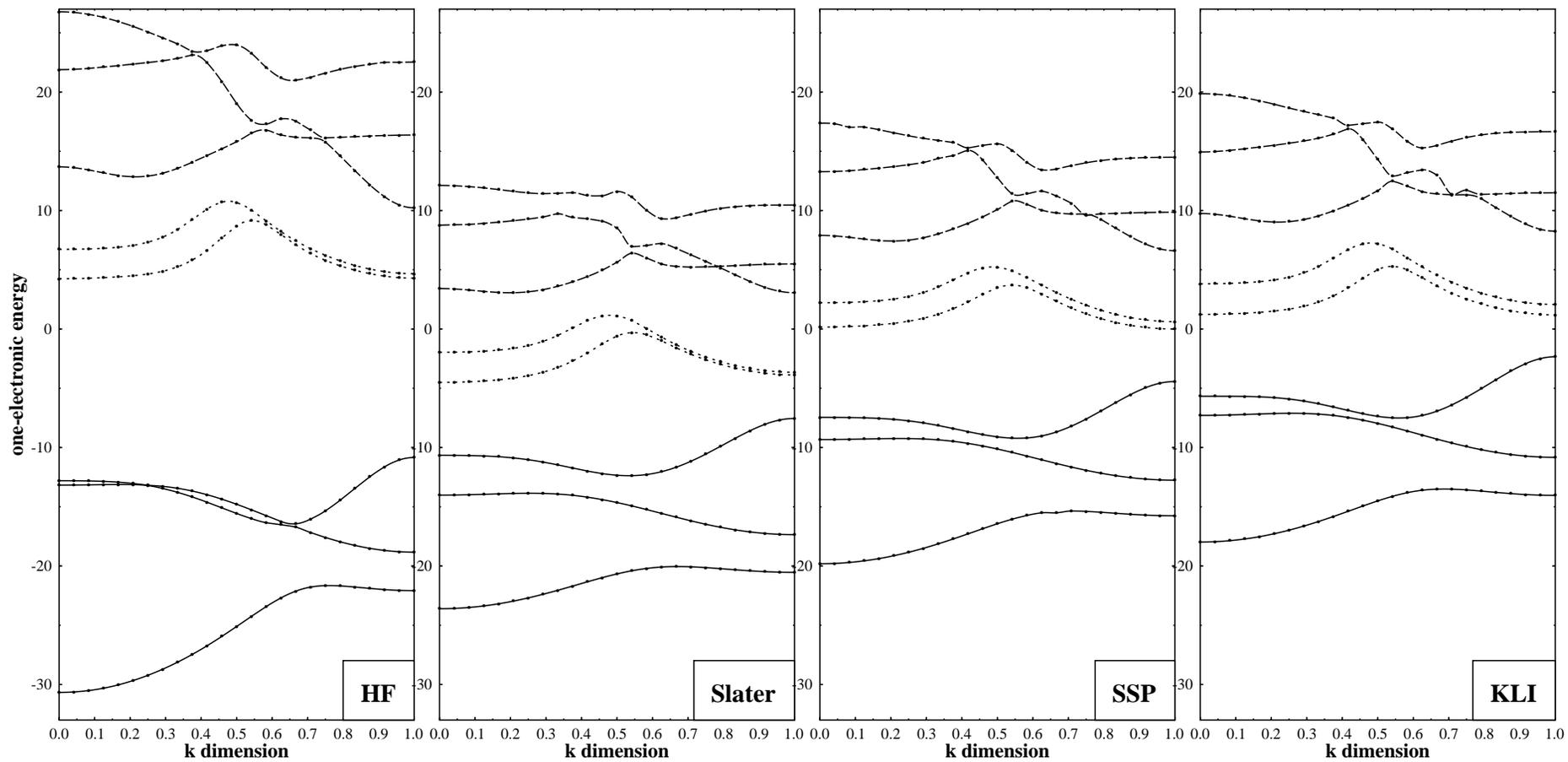
\end{document}